\documentclass[final,5p,times,twocolumn]{elsarticle}

\usepackage{amssymb}
\usepackage{amsmath}
\usepackage{xcolor}
\usepackage[normalem]{ulem} 

\begin{document}

\begin{frontmatter}



\title{Heterogeneous noise-induced extreme events and synchronization in a globally coupled network of FitzHugh-Nagumo oscillators}


\author[inst1]{S. Hariharan}

\affiliation[inst1]{organization={Centre for Nonlinear Science \& Engineering, Department of Physics, SEEE, SASTRA Deemed to be University},
	addressline={}, 
	city={Thanjavur},
	postcode={613401}, 
	state={TamilNadu},
	country={India}}

\author[inst1]{R. Suresh}
\ead{suresh@eee.sastra.edu}
\author[inst1]{V. K. Chandrasekar}

\begin{abstract}
 This study investigates the dynamics of a globally coupled network of heterogeneous FitzHugh–Nagumo (FHN) oscillators under stochastic influences, with particular emphasis on the emergence of extreme events (EE). While previous studies explored FHN networks subjected to homogeneous noise, revealing behaviors such as noise-induced synchronization, stochastic resonance, and coherence resonance, the impact of noise heterogeneity remains poorly understood. Moreover, the emergence of EE under heterogeneous stochastic excitation has largely been overlooked. To address these gaps, we capture the natural variability in neuronal responses to external stimuli by introducing non-identical noise sources, thereby reflecting diversity across the network. Our results reveal that EE can arise both globally, where large excursions occur collectively across the entire network, and partially, where only a subset of oscillators exhibits extreme activity depending on the interplay between noise intensity and coupling strength. We further identify three distinct classes of EE that enrich the system’s dynamical repertoire and propose a quantitative metric capable of distinguishing between global and local occurrences. Remarkably, we demonstrate that even under heterogeneous noise inputs, noise can synchronize EE across the network, underscoring the robustness of collective dynamics in stochastic regimes. Furthermore, causal interaction analysis between oscillator pairs provides mechanistic insights into the initiation and propagation of EE. To the best of our knowledge, this constitutes the first demonstration of both partially and globally synchronized EE triggered solely by noise in a network of coupled oscillators. These findings enhance our understanding of noise-driven collective behavior in complex systems and provide new insights into neuronal dynamics under random influences.
\end{abstract}

\end{frontmatter}

\section{Introduction}
\label{sec1}
Extreme events (EE) are rare, large-amplitude deviations that significantly depart from a system’s typical behavior and are often associated with catastrophic consequences. Their manifestations are wide-ranging, including rogue waves in oceans \cite{chabchoub2011rogue}, tsunamis triggered by seismic activity \cite{sura2011general}, solar flares, epileptic seizures \cite{lehnertz2006}, power-grid failures \cite{dobson2007}, structural collapses \cite{suresh2021}, and financial crashes \cite{bunde2012crash, helbing2001social}. These examples highlight the interdisciplinary significance of EE and the pressing need to identify its mechanisms of origin \cite{albeverio2006extreme}.

Due to their rarity and unpredictability, empirical data from real-world systems alone are often insufficient for understanding EE. As a result, researchers rely heavily on nonlinear dynamical models, whose complex behavior captures the intermittent and extreme responses observed in real systems. Several oscillator-based models, including the Duffing, Li\'enard, Hindmarsh--Rose (HR), and FitzHugh--Nagumo (FHN) systems, have been employed to investigate EE, uncovering multiple generation routes such as crisis-induced intermittency, attractor switching, and bifurcation-driven instabilities \cite{chowdhury2022review}. In neuronal systems in particular, HR and FHN models have provided critical insights into seizure-like bursts~\cite{boaretto2025noise} and dragon-king-type extremes~\cite{mishra2018hr, vijay2023hr, olenin2023hr}. Likewise, discrete-time oscillator networks have revealed intermittent synchronization associated with EE~\cite{kanagaraj2024unraveling}. Despite these advances, a comprehensive understanding of how stochastic influences govern the emergence of EE remains incomplete.

An important but less explored dimension of EE is the role of noise. Neuronal systems are inherently noisy, with variability arising from synaptic transmission, ion channel gating, and environmental fluctuations. Far from being a mere disturbance, noise can facilitate order in excitable systems \cite{faisal2008noise}. It has been shown to enhance weak signals through stochastic resonance \cite{longtin1997sr}, generate coherent oscillations in otherwise quiescent systems via coherence resonance \cite{pikovsky1997cr}, and induce novel rhythms through self-induced stochastic resonance \cite{muratov2005sisr}. In coupled neuronal systems, noise can promote synchronization and pattern formation \cite{wang2000coherence, tessone2004synchronization, ciszak2003anticipating}, even between units that are not directly coupled \cite{zambrano2010synchronization}. Zaks et al. \cite{zaks2005noise} demonstrated that Gaussian noise in coupled FHN systems can induce transitions from subthreshold oscillations to spiking activity. Notably, noise-induced synchronization in the FHN model has also been investigated in the anti-resonance regime, providing insights relevant to deep-brain stimulation in Parkinson’s disease~\cite{touboul2020noise}. More recently, Jinjie \textit{et al.}~\cite{zhu2025complex} reported that an FHN network operating in the self-induced stochastic resonance (SISR) regime exhibits complex synchronization patterns, further confirming the organizing role of noise. Collectively, these studies establish noise as a fundamental driver of emergent dynamics in excitable networks.

Within the context of EE, noise can act as a trigger for rare excursions across stability boundaries, amplifying subthreshold oscillations into extreme spikes. Prior studies have demonstrated EE in excitable systems both in deterministic regimes \cite{feudel2013fhn, feudel2014fhn,feudel2017delay} and under stochastic perturbations \cite{ zaks2005noise,hariharan2025}. For instance, Ansmann et al. \cite{feudel2013fhn} and Karnatak et al. \cite{feudel2014fhn} showed that EE in coupled excitable oscillators can arise via crisis-induced intermittency and attractor switching. More recently, Boaretto \textit{et al.}~\cite{boaretto2025noise} reported noise-induced EE in a Hodgkin-Huxley neuronal network, emphasizing that stochasticity can serve as both a destabilizing and an organizing influence in neuronal ensembles. These works collectively underscore the pivotal role of noise in shaping EE dynamics.

The FHN oscillator, a simplified form of the Hodgkin--Huxley model \cite{fitzhugh1961impulses}, has been widely used to study excitability under different network architectures. The model’s reduced complexity allows it to capture the essential spiking behavior of neurons while remaining analytically tractable. In contrast, the Morris-Lecar model is conductance-based~\cite{morris1981voltage}, and the Hindmarsh--Rose model can reproduce bursting and chaotic dynamics~\cite{hindmarsh1984model}. Since this work focuses on the excitability aspect of neuronal dynamics, the FHN model provides a suitable and computationally efficient framework for exploring EE. Previous works examined seizure-like activity in small-world FHN networks \cite{gerster2020}, the role of pacemaker hubs in shaping collective rhythms \cite{scialla2021}, stabilization effects of adaptive coupling \cite{plotnikov2016}, and multistability in multilayer networks \cite{gorjao2018}. Further studies reported that intermittent synchronization can precede EE in small-world FHN networks \cite{cornejo2024}. That unidirectional inter-layer coupling can promote EE propagation in multiplex architectures \cite{shashangan2025}, with potential strategies to mitigate such events \cite{shashangan2024}. Despite this progress, the essential role of noise, especially heterogeneous noise sources across oscillators, remains largely overlooked in these models, even though biological neuronal systems naturally exhibit non-identical fluctuations \cite{faisal2008noise,white2000channel}.

Motivated by these gaps, the present work investigates EE in a globally coupled network of FHN oscillators subjected to non-identical noise sources, thereby reflecting the intrinsic variability of neuronal populations. Extending our earlier study on noise-induced EE in a single oscillator \cite{hariharan2025}, we generalize the analysis to an ensemble where heterogeneity and coupling jointly shape collective dynamics. The objectives of this study are threefold: (i) to systematically explore how noise heterogeneity across oscillators gives rise to different classes of EE, distinguishing localized events confined to a subset of oscillators from globally coherent EE, thereby providing a comprehensive classification of dynamical states; (ii) to introduce and validate a novel quantitative measure that reliably differentiates between local and global EE regimes, while also distinguishing EE from non-extreme spiking behaviors, enabling the construction of detailed phase diagrams and identification of critical parameter domains; and (iii) to investigate how noise and coupling strength interact to facilitate or regulate EE, with particular focus on whether noise enhances or diminishes coherent activity and the conditions under which transitions between different EE regimes emerge. 

Our analysis reveals five distinct dynamical regimes: quiescent states, localized EE, partially coherent EE, globally coherent EE, and noise-driven non-extreme spiking states, and demonstrates that synchronized EE can robustly emerge in heterogeneous oscillator networks purely under stochastic forcing, even in the absence of identical inputs. Using the Hilbert transform to extract phase information and computing the Kuramoto order parameter \cite{rosenblum1996phase, rosenblum2001phase}, we show that noise-induced synchronization can organize EE across the network, leading to coherent collective dynamics despite heterogeneity. To our knowledge, this is the first evidence of partially and globally synchronized EE triggered by non-identical noise sources in excitable networks. These findings advance the understanding of how stochasticity, heterogeneity, and coupling interplay to generate extreme collective dynamics, offering new perspectives for predicting and controlling EE in neuronal and other complex systems.

\section{Network of FHN oscillators}
\label{sec2}
The mathematical model describing a globally coupled network of FHN oscillators influenced by heterogeneous noise is given by:
\begin{eqnarray}
\label{eq:e2}
\dot{x_i} &=& x_i(a-x_i)(x_i-1)-y_i+\sqrt{2D}\xi_i(t)+\frac{\kappa}{N}\sum_{j=1}^{N}(x_j-x_i),\nonumber\\ 
\dot{y_i} &=& \epsilon(x_i-cy_i),
\end{eqnarray} 

where $i = 1, 2, \dots, N$. The parameters $a$, $\epsilon$, and $c$ specify the intrinsic dynamics of each oscillator, $D$ denotes the global noise intensity, and $\kappa$ represents the strength of global coupling. To incorporate heterogeneity among the noise sources, we assign each oscillator a distinct standard deviation defined as $\sigma_i = i \times 0.01$. This ensures that the noise acting on each oscillator is drawn from a Gaussian distribution with a unique variance with zero mean, thereby generating non-identical stochastic perturbations that evolve independently over time. These noise processes satisfy $\langle \xi_i(t) \rangle = 0$ and $\langle \xi_i(t)\xi_j(s) \rangle = \delta_{ij}\delta(t-s)$, ensuring both temporal and spatial statistical independence. Here, $\delta_{ij}$ is the Kronecker delta and $\delta(t-s)$ is the Dirac delta function.

Equation~(\ref{eq:e2}) provides a minimal yet effective framework for modeling the excitable dynamics of neurons under stochastic fluctuations. Unlike many prior works that assume either a deterministic setting or identical noise across all units, our formulation explicitly highlights the influence of noise heterogeneity on collective network behavior, enabling a systematic examination of how variability in stochastic inputs, when combined with coupling, can generate diverse classes of extreme events ranging from localized to fully synchronized phenomena.
\begin{figure}[h!]
\centering
\includegraphics[width=1.0\linewidth]{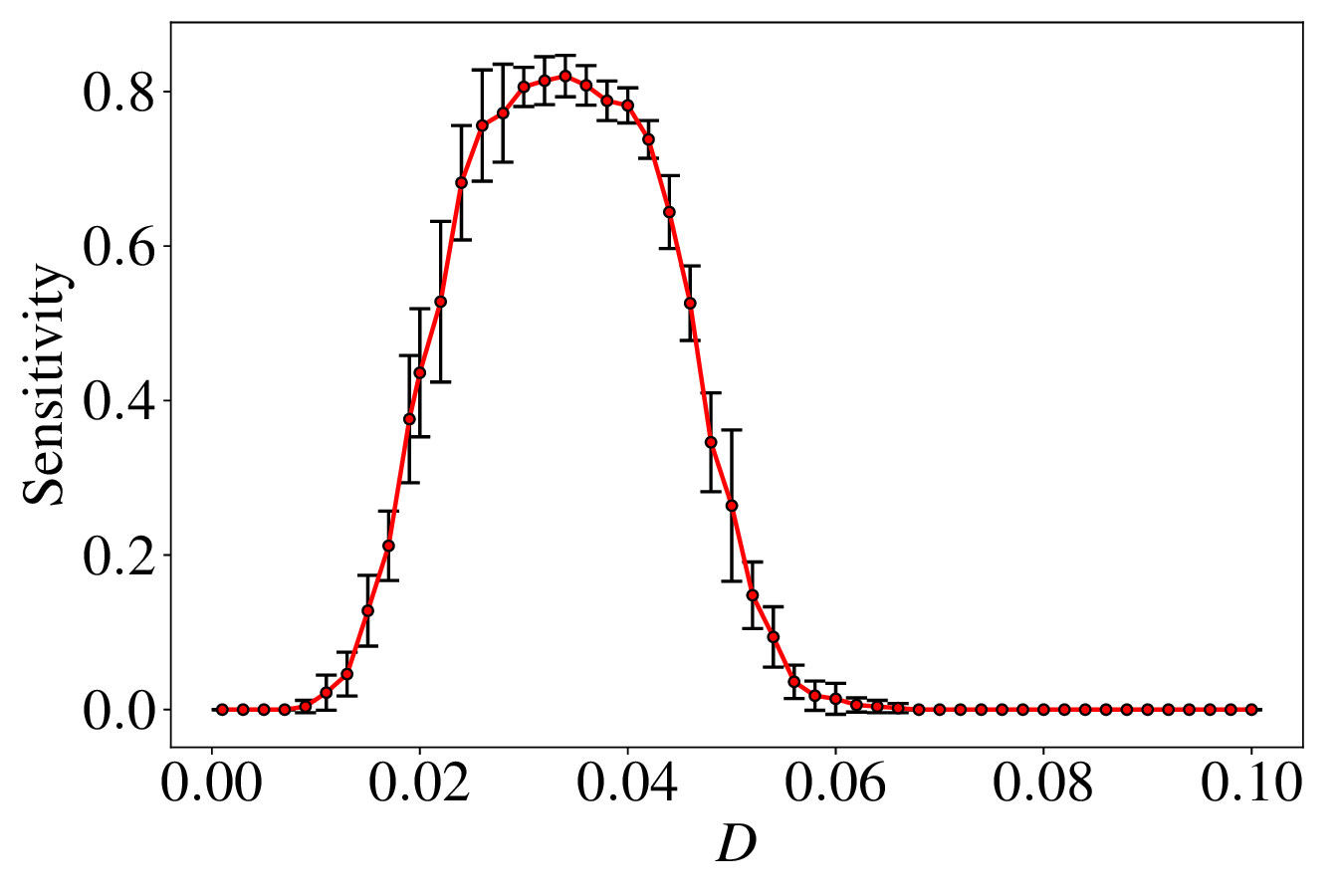}
\caption{Sensitivity analysis of EE with respect to noise intensity $D$. For each value of $D$, the coupling strength $\kappa$ is varied within the range $(0, 0.1)$ using 50 uniform divisions, and the corresponding averaged value is used as the sensitivity measure. Each data point represents the mean of 50 independent realizations, with error bars indicating the standard deviation across simulations.}
\label{gaus}
\end{figure} 

\begin{figure*}[h!]
\centering
\includegraphics[width=1.0\linewidth]{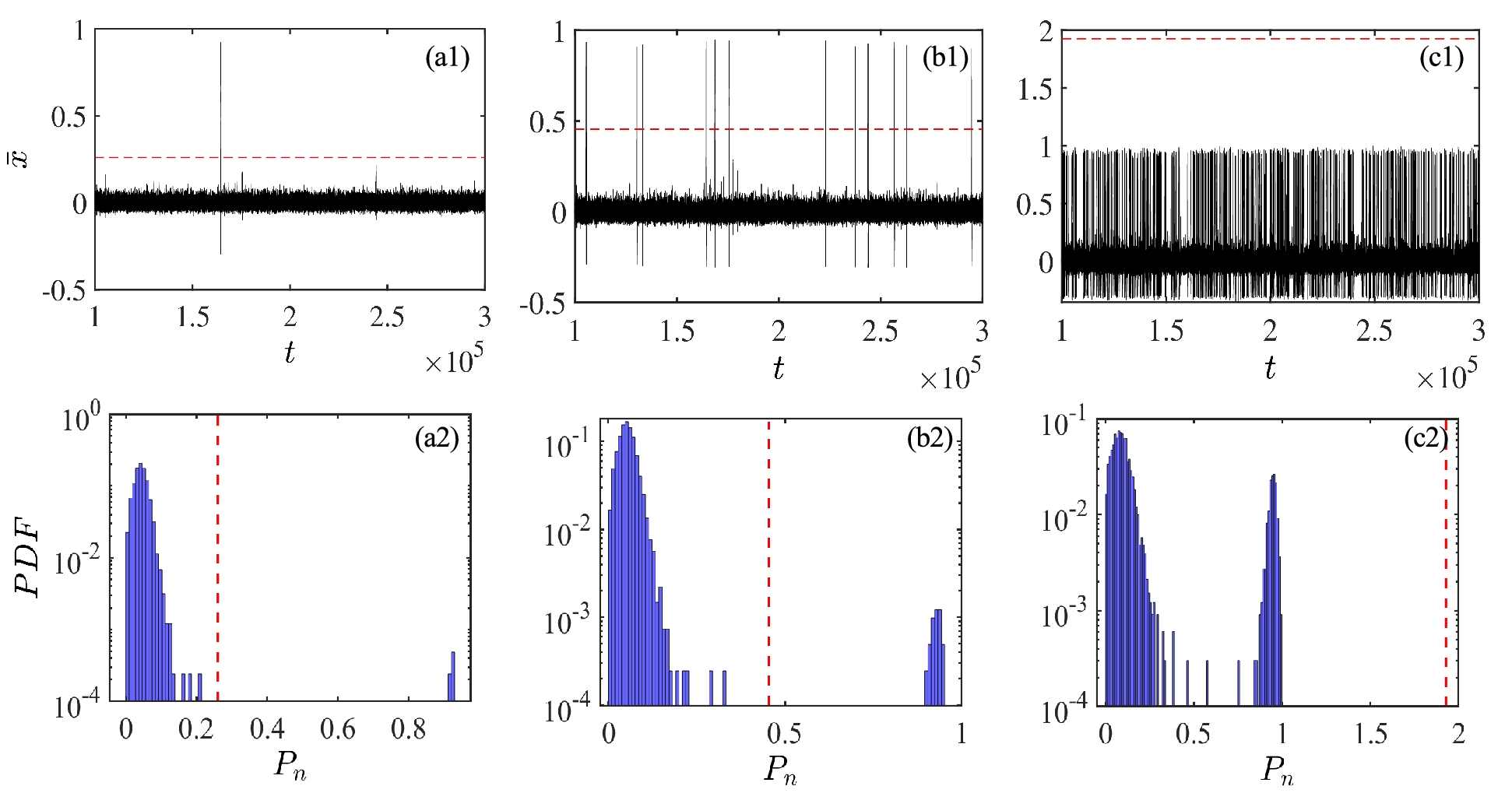}
\caption{\label{t100}Time series and probability density function (PDF) profiles of the global average ($\bar{x}$) for a globally coupled network of FHN oscillators subjected to non-identical noise sources, shown for various noise intensities: (a1–a2) $D$ = 0.02, (b1–b2) $D$ = 0.03, and (c1–c2) $D$ = 0.1. The coupling strength is fixed at $\kappa$ = 0.1. In each subplot, the red dashed horizontal and vertical lines denote the threshold level $H_T$ used to identify EE, as determined from the corresponding time series data.}
\end{figure*}

\begin{figure*}[h!]
\centering
\includegraphics[width=1.0\linewidth]{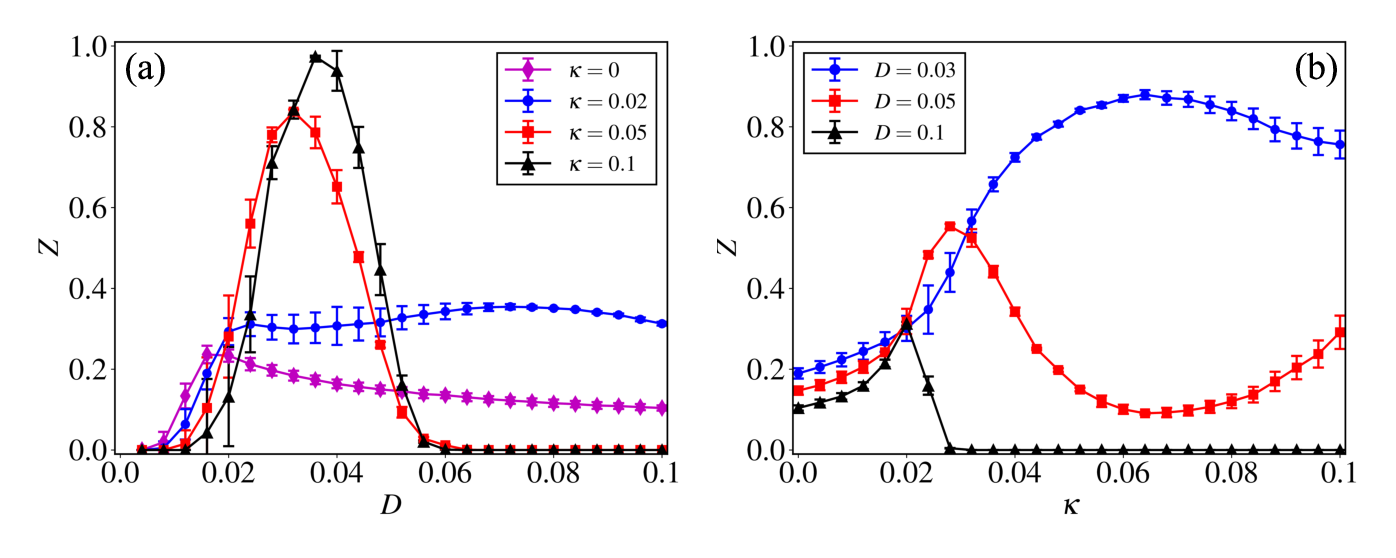}
\caption{\label{2_ee_meas100} Extreme event measure $Z$ for (a) varying coupling strengths $\kappa$ as a function of noise intensity $D$, and (b) varying $\kappa$ for different fixed values of $D$. Each data point represents the mean of 50 independent realizations, with error bars showing the standard deviation across trials. The plots illustrate how the interplay between noise intensity and coupling strength influences the emergence of EE in the network (\ref{eq:e2}).}
\end{figure*}
\begin{figure*}[h!]
\centering
\includegraphics[width=0.55\linewidth]{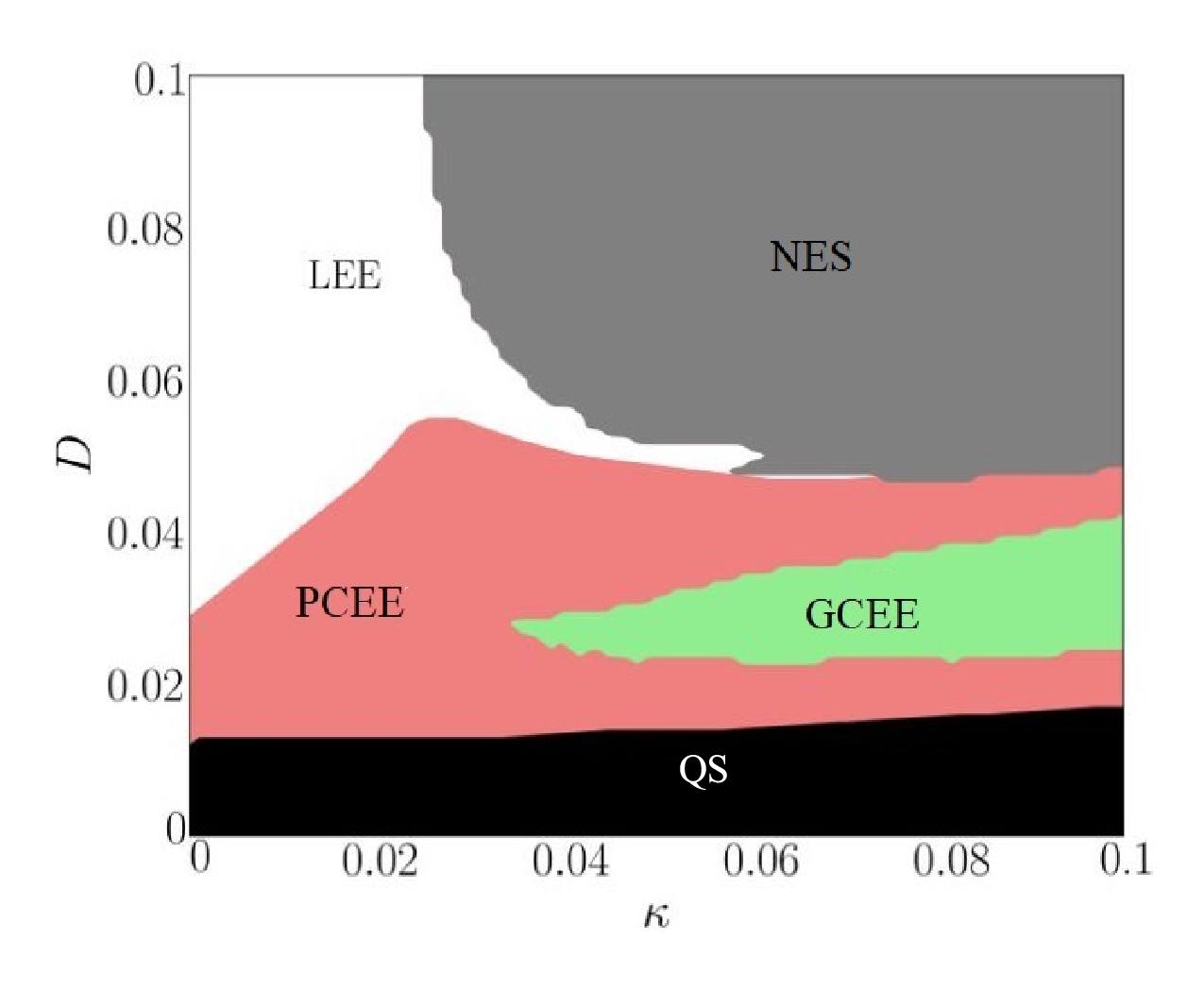}
\caption{\label{2_ee_meas_par100} Two-parameter phase diagram illustrating the emergence of five distinct dynamical states in the $(\kappa, D)$ parameter space, where $\kappa \in (0, 0.1)$ is the coupling strength and $D \in (0, 0.1)$ is the noise intensity. The diagram delineates regions corresponding to: quiescent/steady state (QS, black), partially coherent extreme events (PCEE, red), globally coherent extreme events (GCEE, green), localized extreme events (LEE, white), and non-extreme spiking (NES, grey). The boundaries highlight how the interplay between noise and coupling determines whether EE occurs locally, globally, or is absent altogether.}
\end{figure*}
To investigate the noise-induced dynamics of the system, we numerically integrate Eq.~\ref{eq:e2} using a fourth-order Runge--Kutta method. Unless mentioned otherwise, the system parameters are fixed at $a =$ 0.12, $\epsilon =$ 0.005, and $c =$ 0.5, with the coupling strength set to $\kappa =$ 0.1. We choose $N$ = 100 for our numerical simulation. In the absence of noise ($D =$ 0), the system rapidly relaxes to a stable fixed point after a short transient period \cite{hariharan2025}. As the noise intensity increases, the oscillators first exhibit subthreshold oscillations. With further increase in noise, the dynamics become more intricate, and the system begins to generate rare, recurrent large-amplitude excursions, which we classify as EE. These extreme spikes appear sporadically against the backdrop of small-amplitude oscillations. 

To identify EE, we employ an amplitude-based threshold criterion~\cite{chowdhury2022review}. An event is classified as extreme if its peak amplitude exceeds $H_T = \langle P_n \rangle + n\sigma_H$, where $\langle P_n \rangle$ and $\sigma_H$ represent the mean and standard deviation of the peak heights, respectively. The parameter $n$ is set to 8 to ensure that only statistically rare and significant deviations are identified as EE. This choice is guided by the requirement that peaks beyond this threshold occur with an exceedingly low probability under the background distribution, thereby ensuring that the detected events genuinely reflect the statistical nature of extremes. 

Since the two primary control parameters governing EE are the noise intensity $D$ and the coupling strength $\kappa$, we perform a sensitivity analysis to quantify the dependence of EE occurrence on these parameters. For each $D \in (0.001, 0.1)$, the coupling strength is varied over $(0, 0.1)$ in 50 divisions, and the averaged value of the collective mean variable $\bar{x}$ crossing the threshold $H_T$ is used as a representative indicator of EE. The results, illustrated in Fig.~\ref{gaus}, show that each data point corresponds to an average over 50 independent realizations, with error bars representing the standard deviation across runs. It is evident that the system exhibits the highest sensitivity to EE formation for $D \in (0.017, 0.058)$ across most coupling strengths. In particular, the range $0.02 \leq D \leq 0.05$ yields the most frequent EE, providing a preliminary indication of the noise intensities that promote sporadic extreme dynamics.
\begin{figure}[h!]
\centering
\includegraphics[width=1.\linewidth]{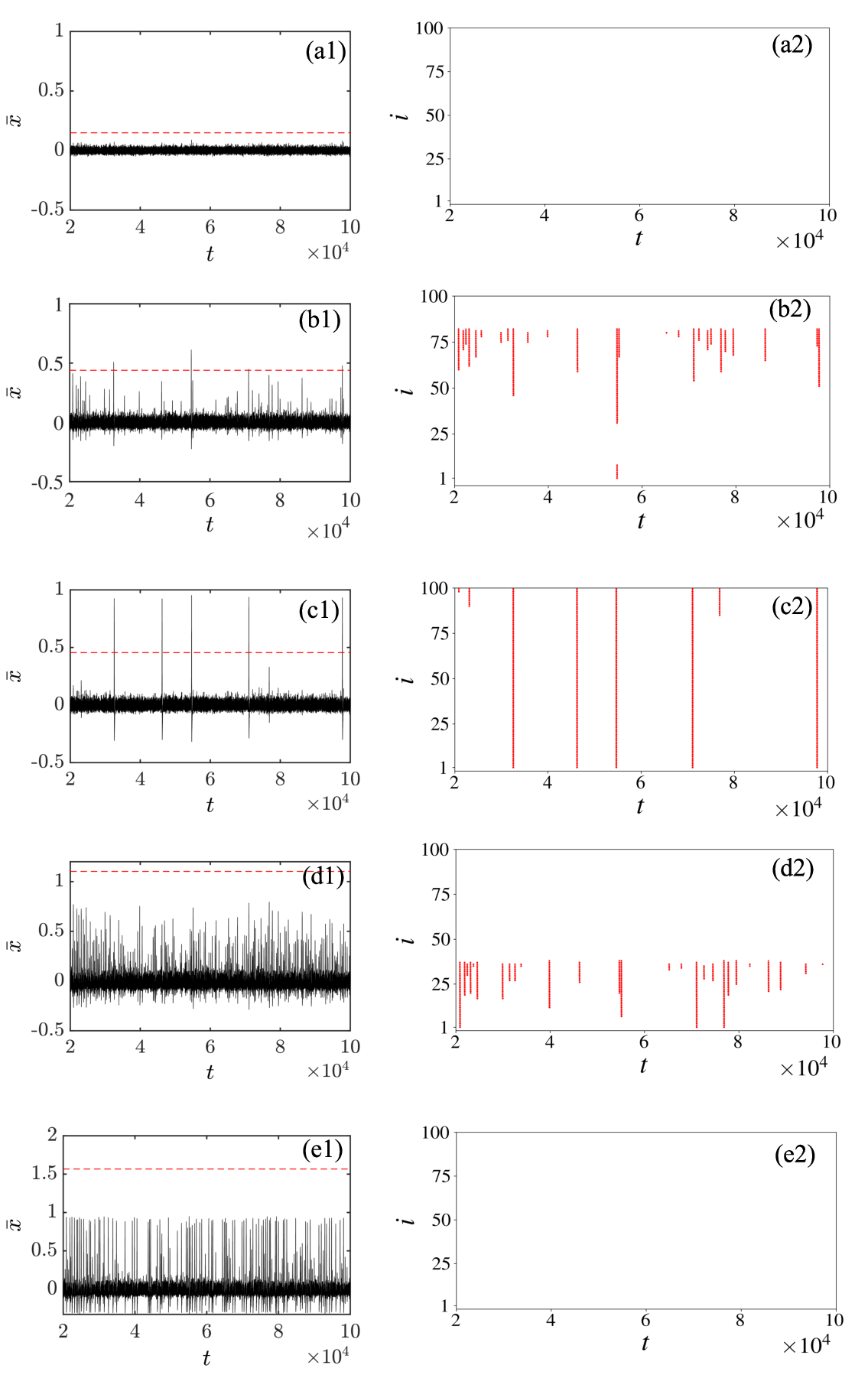}
\caption{\label{2_t}Distinct dynamical states of the oscillator network for different combinations of coupling strength ($\kappa$) and noise intensity ($D$). Each row shows (left) the time series of the global average variable $\bar{x}$, where the dashed red line denotes the EE threshold $H_T$, and (right) the corresponding spatiotemporal plot indicating oscillators whose amplitudes exceed $H_T$. The panels illustrate: (a) Quiescent state (QS): subthreshold oscillations or near-steady behavior ($\kappa = 0.05$, $D = 0.008$); (b) Partially coherent EE (PCEE): intermittent EE visible both at the individual and global levels ($\kappa = 0.02$, $D = 0.03$); (c) Globally coherent EE (GCEE): large, synchronized EE across all oscillators ($\kappa = 0.1$, $D = 0.03$); (d) Localized extreme events (LEE): EE confined to subsets of oscillators without global excursions ($\kappa = 0.02$, $D = 0.08$); and (e) Non-extreme spiking (NES): frequent large-amplitude oscillations that fail to exceed $H_T$ ($\kappa = 0.05$, $D = 0.08$). Panels (a2) and (e2) appear blank because they correspond to the QS and NES regimes, respectively, where no oscillator surpasses the predefined extreme-event threshold $H_T$; hence, no activity is displayed in the spatiotemporal plots.}
\end{figure}

To visualize the system’s dynamics, we present in Fig.~\ref{t100} the time evolution of the mean variable $\bar{x}= \frac{1}{N} \sum_{i=1}^{N} x_i$, which reflects the collective behavior of all $N$ oscillators. To characterize the nature of the observed large-amplitude oscillations, we also compute the corresponding probability density functions (PDF) for varying noise intensities. Figure~\ref{t100}(a1) shows that for a noise intensity of $D = 0.02$, a single large spike emerges amidst small-amplitude oscillations. The red dashed horizontal line indicates the threshold height $H_T$, and the extreme spike exceeds this value, thereby qualifying as an EE. The corresponding PDF in Fig.~\ref{t100}(a2), computed over a long simulation time of $5 \times 10^9$ time units, exhibits a heavy-tailed distribution with a sharp drop-off near the threshold, confirming the rare nature of the event.

When the noise intensity is increased to $D = 0.03$ [Fig.\ref{t100}(b1–b2)], multiple large-amplitude events appear in the time series. The associated PDF reveals a higher likelihood of these events, indicating an increase in the frequency of EE. At higher noise levels ($D = 0.1$,  [Fig.\ref{t100}(c1–c2)]), the spikes become frequent and quasi-regular, blurring the distinction between typical oscillations and rare EE. The PDF broadens significantly, indicating a transition to dynamics dominated by recurrent large-amplitude oscillations rather than isolated extremes.

These observations suggest that a critical range of noise intensity exists where EE emerges as rare, large excursions. Beyond this range, the system transitions into a regime dominated by frequent large oscillations, making the identification of true EE less meaningful. 

So far, we have investigated the system’s behavior from a collective standpoint by analyzing the global average of the dynamical variable, $\bar{x}$. This approach has allowed us to detect the emergence of EE as a function of noise intensity. However, since the oscillators in the network are subjected to heterogeneous noise sources, it becomes essential to explore how these EE are distributed across individual units. Specifically, we aim to determine whether EE is localized, i.e., restricted to a few individual oscillators, or whether it reflects a coordinated, system-wide phenomenon shaped by the interplay of noise heterogeneity and coupling strength. To this end, we now shift our focus to the local oscillator dynamics and develop a quantitative framework to characterize the spatial extent of EE within the network.
\subsection{Quantifying the spatial extent of extreme events}
To distinguish whether EE occurs partially, where only a subset of oscillators are active, or globally, in which all oscillators participate, we define a quantitative measure based on a threshold criterion. For each oscillator, the maximum membrane potential $x_{i_{\mathrm{max}}}$ during an event is compared against a predefined threshold $H_{T_i}$ using the Heaviside function:
\begin{equation}
X_i = \Theta(x) \Rightarrow
\begin{cases} 
	1, & \text{if } x_{i_{\text{max}}} > H_{T_{\text{i}}} \\
	0, & \text{otherwise,} 
\end{cases}
\end{equation}
where $X_i$ = 1 indicates that the $i-th$ oscillator exhibits EE. Averaging over all $N$ oscillators yields:
\begin{equation}
Z = \frac{1}{N} \sum_{i=1}^{N} X_i
\label{z}
\end{equation}
which quantifies the spatial extent of EE activity within the network. By definition, $Z = 1$ corresponds to a globally coherent EE, where all oscillators are simultaneously active, while $Z = 0$ indicates a complete absence of EE. Intermediate values $0 < Z < 1$ reflect partial EE states, where only a fraction of oscillators participate. Although $Z = 1$ theoretically represents perfect global coherence, averaging over 50 independent realizations introduces variability that prevents $Z$ from reaching exactly one. Instead, the error bars typically span values close to unity, with the mean $Z$ stabilizing around 0.8--0.9. Therefore, we adopt $Z > 0.8$ as a practical and statistically robust criterion for identifying global EE, effectively capturing collective behavior while accounting for stochastic fluctuations.

Figures~\ref{2_ee_meas100}(a) and \ref{2_ee_meas100}(b) provide quantitative insights into the emergence and prevalence of EE. Fig. \ref{2_ee_meas100}(a) presents a one-parameter diagram illustrating how the fraction of oscillators exhibiting EE, quantified by the measure $Z$, varies with noise intensity $D$ for different fixed coupling strengths $\kappa$. Each data point represents the mean of 50 independent realizations, with error bars representing the standard deviation across trials. In the absence of coupling ($\blacklozenge$), for $D < 0.012$, all oscillators remain in small-amplitude oscillatory states with $Z \approx 0$. Within the range $0.012 < D < 0.1$, the network enters a partial EE regime. For instance, at $D = 0.02$, approximately 30\% of oscillators exhibit EE, demonstrating that even weak noise can trigger partial EE in an uncoupled network. As $D$ increases further, $Z$ gradually decreases, and no transition to a globally coherent EE state occurs in this case.

When coupling is increased slightly to $\kappa = 0.02$ ($\medbullet$), the fraction of oscillators exhibiting EE rises with $D$ up to $\approx 0.02$, after which $Z$ fluctuates within the range $0.2 < Z < 0.4$ up to $D = 0.1$. This indicates that weak coupling enhances the network’s responsiveness to stochastic perturbations. At intermediate coupling ($\kappa = 0.05$, $\blacksquare$), the system exhibits only small-amplitude oscillations for $D < 0.013$. In the interval $0.013 < D < 0.02$, partial EE emerges, and at $D \approx 0.024$, all oscillators begin to exhibit EE simultaneously, yielding $Z > 0.8$, which we take as a practical indicator of global EE. For $D > 0.05$, collective activity diminishes and EE gradually vanishes, as reflected in the reduction of $Z$.

For stronger coupling ($\kappa = 0.1$, $\blacktriangle$), no EE are observed for $D \leq 0.02$. However, for $D > 0.024$, the system undergoes a rapid transition, first to a partial EE state ($Z \approx 0.5$) and subsequently to global EE within the range $0.025 < D < 0.04$ ($Z > 0.8$). Beyond $D > 0.044$, $Z$ drops sharply, indicating that excessive noise drives frequent large-amplitude spiking activity that fails to exceed the EE threshold $H_T$, thereby inhibiting the occurrence of EE even under strong coupling conditions.

\begin{figure*}[h!]
\centering
\includegraphics[width=1.0\linewidth]{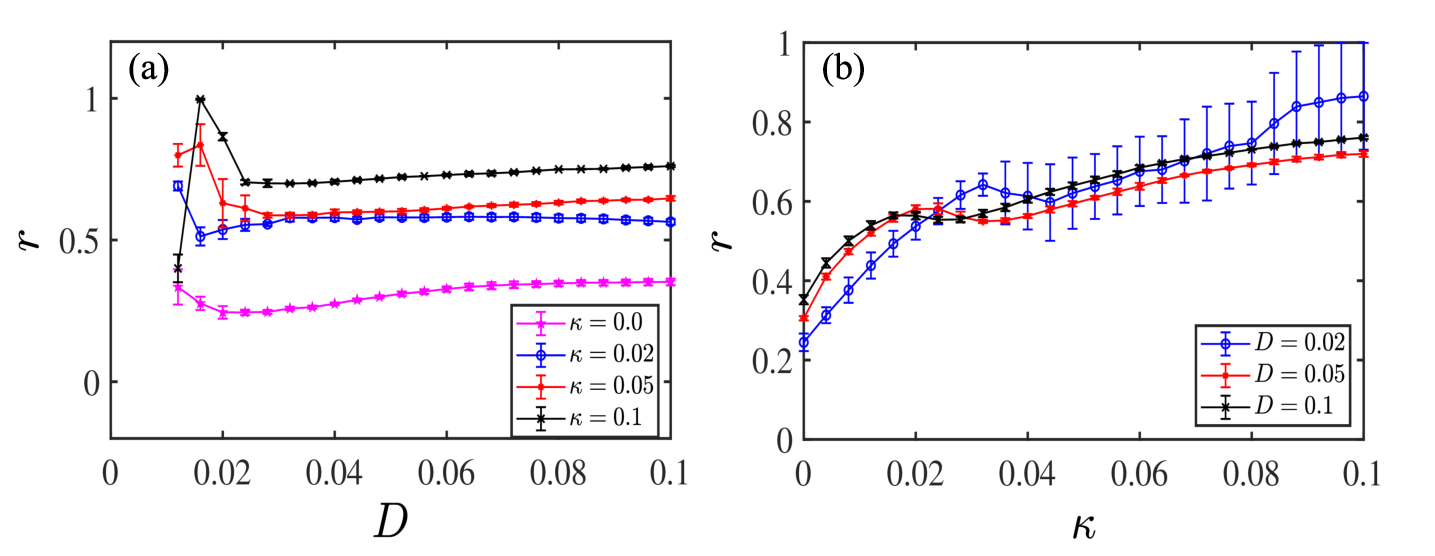}
\caption{\label{op100}(a). Spike-based synchronization analysis using the order parameter $r$, computed by considering only the spike timings of the oscillators. Each data point represents the mean of 50 independent realizations, with error bars showing the standard deviation across trials.	(a) Order parameter $r$ plotted as a function of noise intensity $D$ for different fixed coupling strengths $\kappa$. A dip in $r$ for intermediate $D$ values indicates desynchronization, especially at lower coupling strengths.	(b) Order parameter $r$ as a function of coupling strength $\kappa$ for different fixed noise intensities. Stronger coupling generally promotes spike synchronization, with a gradual increase in $r$ across all noise levels.}
\end{figure*}

\begin{figure}[h!]
\centering
\includegraphics[width=1.0\linewidth]{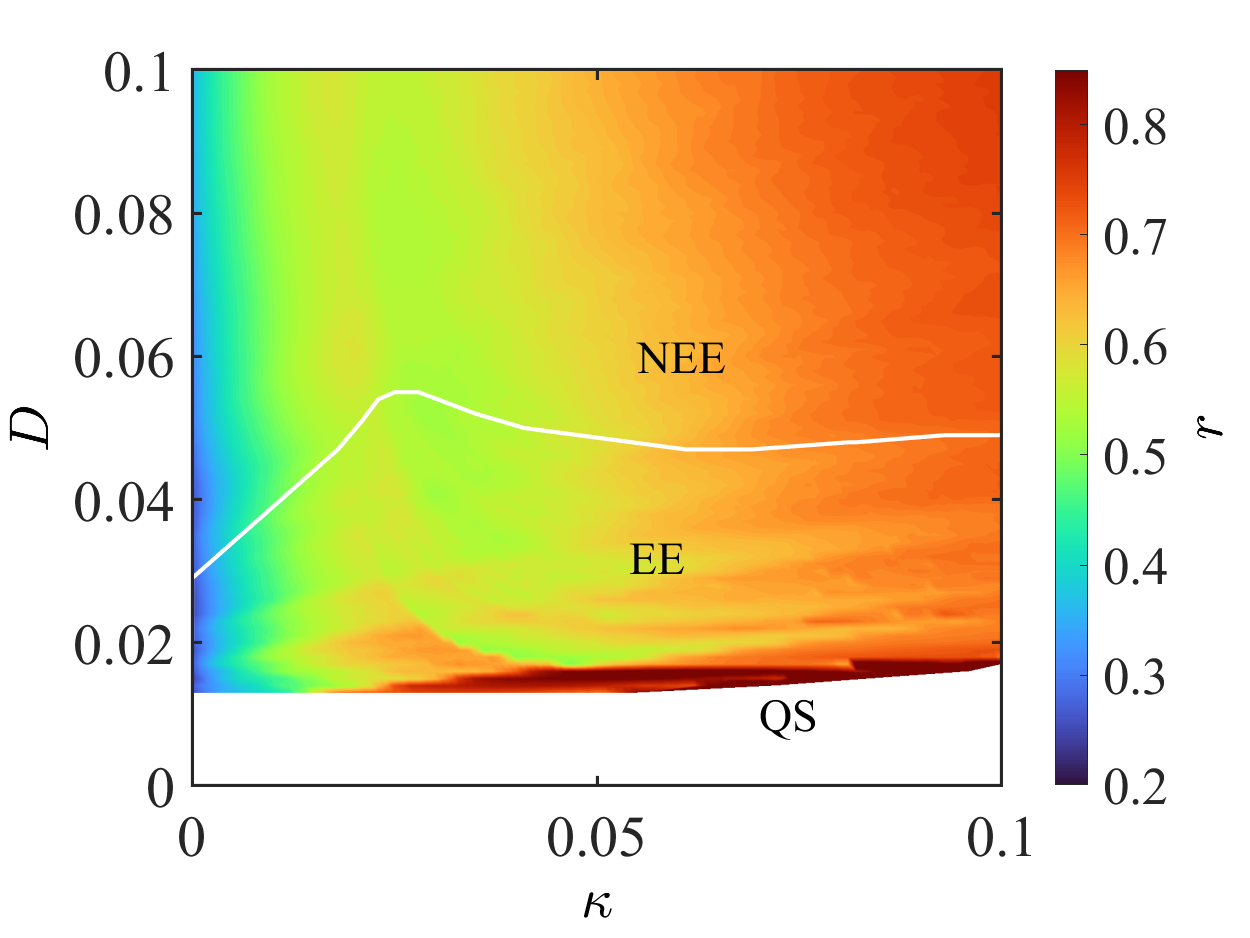}
\caption{\label{2par_op_100} Two-parameter phase diagram in the $(\kappa, D)$ space, illustrating the degree of synchronization of EE based on the spike-timing order parameter $r$, as indicated by the color bar. Higher values of $r$ correspond to greater synchronization among spike events. The closed contour encloses the region where EE is observed, while the white region denotes parameter combinations where no spiking activity occurs (QS).}
\end{figure}
Figure \ref{2_ee_meas100}(b) depicts the variation of $Z$ with coupling strength $\kappa$ for fixed noise intensities $D$, with error bars indicating the standard deviation over 50 independent realizations. For weak noise ($D = 0.03$, $\medbullet$), the network exhibits partial EE for $\kappa \leq 0.032$. Beyond this threshold, a sharp transition to global EE occurs, with $Z \gtrsim 0.8$ sustained across higher coupling values. This indicates that under weak noise, increasing coupling effectively promotes network-wide EE. For moderate noise ($D = 0.05$, $\blacksquare$), the system initially displays weak partial EE at $\kappa < 0.02$. As $\kappa$ increases within $0.02 < \kappa < 0.036$, the fraction of oscillators exhibiting EE rises to approximately 50\%, but a transition to global EE does not occur. For $\kappa \gtrsim 0.04$, $Z$ sharply declines toward zero, signifying the absence of both partial and global EE. At higher noise intensity ($D = 0.1$, $\blacktriangle$), about 20\% of oscillators exhibit partial EE for $\kappa \leq 0.02$, but EE vanish completely once $\kappa$ exceeds 0.02. These observations confirm that coupling can facilitate EE under weak noise, whereas stronger noise disrupts coherence and suppresses both partial and global EE at higher coupling strengths.

Overall, the results reveal a delicate interplay between noise intensity and coupling strength in determining the emergence of EE. Weak noise combined with moderate coupling favors global EE, while excessive noise or overly strong coupling inhibits both partial and global EE. Moderate coupling thus appears optimal for inducing EE under weak-to-intermediate noise conditions, whereas high noise intensity prevents EE regardless of coupling strength.

While $Z$ robustly distinguishes the spatial extent of EE, it does not capture the temporal synchrony or clarify the distinction between intermediate states. To improve classification, we introduce an additional criterion based on the time-averaged global activity $\bar{x}$. When $0 < Z < 1$ and $\bar{x} > H_T$, the system is classified as exhibiting partially coherent EE, indicating substantial but incomplete synchrony among oscillators. Conversely, when $0 < Z < 1$ and $\bar{x} < H_T$, the EE are localized, with only a few oscillators spiking in isolation.

For cases where $Z = 0$, both quiescent and non-extreme spiking states are possible. To separate these, we use the criterion that $\bar{x} > 0.1$ signifies a non-extreme spiking state, characterized by frequent, subthreshold excitations. In contrast, $\bar{x} < 0.1$ indicates quiescent states, a low-activity, noise-driven regime. The threshold $\bar{x}=0.1$ is chosen based on observed dynamics: quiescent dynamics remain confined to low-amplitude fluctuations ( $\bar{x}< $0.1, Fig. \ref{2_t}(a1), whereas non-extreme spiking exhibits frequent noise-driven spiking that elevates the global average above this value (Fig. \ref{2_t}(e1)).

By applying these criteria, we map out five distinct dynamical states in the network:
\begin{enumerate}
\item \textbf{Quiescent state (QS):} Absence of EE, with both individual and global activity below threshold;
\item \textbf{Partially coherent extreme events (PCEE):} A subgroup of oscillators generates temporally correlated EE, lifting $\bar{x}$ above threshold but without complete network-wide synchronization;
\item \textbf{Globally coherent extreme events (GCEE):} Near-simultaneous excitation across the entire network, with strong collective response ($Z > 0.8$);
\item \textbf{Localized extreme events (LEE):} Sporadic and spatially sparse excitation in a few oscillators, insufficient to affect the global average;
\item \textbf{Non-extreme spiking (NES):} Frequent high-amplitude oscillations, with no excursions surpassing the EE threshold either locally or globally.
\end{enumerate}

Figure~\ref{2_ee_meas_par100} presents the two-parameter phase diagram in the $(\kappa, D)$ plane. Network states are classified using the measure $Z$ in conjunction with the additional criteria described above. The diagram reveals five distinct dynamical regimes, each corresponding to qualitatively different EE behaviors, and these are color-coded for clarity. Representative examples of each regime are shown in Fig.~\ref{2_t}, which includes time series of $\bar{x}$ and spatiotemporal activity plots showing the individual oscillator dynamics as a function of time $t$. In the time series, the horizontal dashed line marks the threshold $H_T$, while red points in the spatiotemporal plots indicate oscillators whose amplitudes exceed $H_T$.

The black region in Fig.\ref{2_ee_meas_par100} corresponds to QS, where the system exhibits small-amplitude oscillations or converges to a steady state, with $Z = 0$. This state occurs only at very low noise intensities ($D < 0.012$), independent of coupling strength. For example, in Figs. \ref{2_t}(a1) and \ref{2_t}(a2) for $\kappa = 0.05$ and $D = 0.008$, the time series shows only minor fluctuations around a fixed point, and the spatiotemporal plot confirms low-amplitude behavior across all oscillators.

The pale red region corresponds to the PCEE state, where EE are present both in the global average and among a significant subset of oscillators. Here, $0 < Z < 0.8$, indicating substantial but incomplete spatial participation. As illustrated in Figs.~\ref{2_t}(b1) and \ref{2_t}(b2) for $\kappa = 0.02$ and $D = 0.03$, the time series of $\bar{x}$ exhibits intermittent excursions above $H_T$, and the spatiotemporal plot shows that roughly 70 out of 100 oscillators spike during EE episodes.

The light green region in Fig.\ref{2_ee_meas_par100}, labeled GCEE, represents globally coherent EE involving nearly all oscillators, with $0.9 < Z \leq 1$. In Figs. \ref{2_t}(c1) and \ref{2_t}(c2) for $\kappa = 0.1$ and $D = 0.03$, the time series clearly shows recurrent, large-amplitude excursions, while the spatiotemporal plot confirms near-complete synchrony during EE episodes. This regime arises from an optimal interplay between noise and coupling.

The white region corresponds to the LEE state, characterized by isolated and sparse EE in individual oscillators without global coherence. This regime occurs at high noise intensities and low-to-moderate coupling, with $0 < Z < 0.8$. For example, in Figs.~\ref{2_t}(d1) and \ref{2_t}(d2) for $\kappa = 0.02$ and $D = 0.08$, the time series of $\bar{x}$ remains below $H_T$, but the spatiotemporal plot shows scattered large spikes among a fraction of individual oscillators.

The grey region represents the NES state, characterized by high-frequency spiking induced by strong noise. Although the network exhibits elevated excitation, the rapid and frequent spiking prevents the formation of rare, large excursions required for EE classification, resulting in $Z = 0$. This behavior is shown in Figs.~\ref{2_t}(e1) and \ref{2_t}(e2) for $\kappa = 0.05$ and $D = 0.08$, where dense spiking activity is present but no large peaks above $H_T$ occur. This behavior can be attributed to the higher levels of noise, which generate a greater number of spikes in an irregular manner. Such behavior supports the findings reported for the single FHN neuron~\cite{hariharan2025}, where, beyond a certain noise intensity, the system exhibits frequent spiking oscillations. Furthermore, similar observations have been made in networked systems, where elevated noise levels lead to the appearance of irregular spike trains~\cite{baspinar2021coherence}. Experimental evidence corroborates these dynamics: neuronal culture studies using optical or electrical stimulation reveal that optimal noise promotes regularity, while excessive noise leads to highly frequent but irregular spikes \cite{kim2015coherence}.

Together, Figs.~\ref{2_ee_meas_par100} and \ref{2_t} provide a comprehensive view of how noise intensity $D$ and coupling strength $\kappa$ determine the emergence and classification of EE states in the network. The transition from QS to GCEE follows a structured sequence through intermediate PCEE or LEE states, indicating that both noise and coupling play crucial roles in this process. Weak noise favors QS, moderate noise can induce either PCEE or GCEE depending on the coupling, and strong noise leads to NES or LEE. Thus, large-scale, coherent EE emerges most prominently within an intermediate range of both parameters.

\subsection{Synchronization in a network of globally coupled oscillators}

After establishing the emergence of distinct types of EE due to heterogeneous noise sources, we now turn our attention to the synchronization dynamics within the network. Although the oscillators are identical in structure, the presence of noise prevents the system from achieving complete synchronization. Therefore, we focus on phase synchronization, as the stochastic nature of the system inherently disrupts amplitude coherence.

To quantify synchronization, we employ the classical Kuramoto order parameter $r$, a widely used metric for assessing phase coherence among oscillators. However, due to the noisy environment, conventional phase extraction methods may not be robust. To address this, we utilize the Hilbert transform approach to extract the instantaneous phase $\phi$ of each oscillator. This method involves computing the analytic signal through the Fourier transform (FT) and inverse FT, from which the instantaneous phase is obtained as:

\begin{eqnarray}
\phi(t) = \tan^{-1}\left(\frac{x_a(t)}{x(t)}\right),
\end{eqnarray}
where $x(t)$ is the original time series, and $x_a(t)$ is the imaginary component of the analytic signal derived via the Hilbert transform. The global synchronization measure is then computed as:

\begin{equation}
r = \left\langle \left| \frac{1}{N}\sum_{i=1}^{N} e^{i\phi_i(t)} \right| \right\rangle_t,
\end{equation}
where $\langle \cdot \rangle_t$ denotes time-averaging. In our analysis, we compute $r$ by considering only the spike times across all $N = $ 100 oscillators when the collective threshold $\bar{x}$ = 0.1 is exceeded.

Figure~\ref{op100}(a) shows the order parameter $r$ as a function of noise intensity $D$ for four different coupling strengths $\kappa = \{0.0, 0.02, 0.05, 0.1\}$. Spike synchronization begins at $D =$ 0.01, where spike activity emerges in the system. For the uncoupled case ($\kappa$ = 0.0,  $\blacklozenge$), a desynchronization with ($r \approx$ 0.35) is observed at $D$ = 0.02, indicating that the systems are uncorrelated due to the lack of coupling. With weak coupling ($\kappa$ = 0.02, $\medbullet$), the system maintains a moderate level of synchronization, though a decline in $r$ is still evident for intermediate noise levels. At intermediate coupling ($\kappa$ = 0.05, $\blacksquare$), the network sustains a higher and more stable partial synchronization level across the noise range with $r\approx$ 0.6. For strong coupling ($\kappa$ = 0.1, $\blacktriangle$), the system maintains the highest synchronization values across all $D$, demonstrating that stronger coupling effectively counteracts the desynchronizing effects of noise. 

Figure \ref{op100}(b) illustrates the variation of the synchronization measure $r$ as a function of the coupling strength $\kappa$ for three fixed noise intensities: $D = 0.03$ ($\medbullet$), $D = 0.05$ ($\blacksquare$), and $D = 0.1$ ($\blacktriangle$). For the lowest noise intensity ($D = 0.03$), the system exhibits relatively high synchronization even at weak coupling, which can be attributed to the presence of sparse, weakly perturbed spiking activity. As the noise intensity increases to $D = 0.05$ and $D = 0.1$, the initial synchronization level at $\kappa \approx 0$ decreases; however, in both cases, $r$ increases monotonically with $\kappa$. This trend confirms that stronger coupling effectively counteracts noise-induced desynchronization, enabling the network to regain a high degree of coherence even under substantial noise perturbations.

To gain a comprehensive understanding of the interplay between noise intensity and coupling strength in governing spike synchronization, we construct a two-parameter phase diagram in the $(D, \kappa)$ space, as shown in Figure~\ref{2par_op_100}. Each point in this diagram represents the average order parameter $r$, calculated from spike timings across all oscillators, with the color scale indicating the degree of synchronization.

This parameter-space landscape reveals three distinct dynamical regimes. The first regime is the QS, which is the complete absence of spiking activity. Second is the EE regime, enclosed by a white contour, which encompasses both PCEE and GCEE, where EE emerges at either partial or global scales. Finally, thenon-extreme spiking Region comprises LLE and NES states, representing conditions in which frequent spike activity occurs without global EE manifestations.

At low noise intensities ($D < 0.017$), the system resides in the QS state. Here, no spiking activity is observed, and the synchronization measure $r$ is either undefined or negligibly small. As noise intensity increases, the system crosses into the EE region, where collective spiking events, classified as PCEE or GCEE, emerge. Synchronization within this region exhibits a non-monotonic dependence on coupling strength. Specifically, at low $D$ and weak coupling ($\kappa <$ 0.02), synchronization is poor, as indicated by small $r$ values. A slight increase in coupling ($\kappa > $ 0.03) yields moderate $r$ values, reflecting partial synchronization or localized spiking, which is consistent with the PCEE state. With further increases in $\kappa$, global synchronization is restored, marking a transition back to GCEE.

In the region above the white contour, corresponding to the non-extreme spiking regime, the global average spike activity becomes more frequent but never attains the threshold height $H_T$ required to qualify as an EE. For low to moderate coupling strengths ($\kappa \in [0.02,0.05]$), the system exhibits weakly synchronized spiking, as indicated by intermediate values of the synchronization measure ($r \in [0.5,0.6]$). With increasing noise intensity $D$ and coupling $\kappa$, synchronization progressively strengthens. 

The dynamical behavior is consistent with the EE classification presented in Figs.~\ref{2_ee_meas100}(a) and \ref{2_ee_meas_par100}, where PCEE is observed at low $D$ evolves into GCEE as $\kappa$ increases. Likewise, within the NEE regime of Fig. \ref{2par_op_100}, increasing both $D$ and $\kappa$ drives the system from partial toward global spike synchronization.

In summary, noise acts as the primary trigger for spike initiation, whereas coupling strength governs the degree of coherence among oscillators. Weak coupling promotes fragmented, localized spiking, whereas strong coupling facilitates globally synchronized EE. Even in the presence of heterogeneous noise, the network exhibits distinct phase-synchronization patterns and smooth transitions between partial and global EE states. These results highlight the delicate balance between stochastic forcing and coupling in shaping the collective dynamics of excitable oscillator networks.

\section{Mechanism of extreme events}
\label{sec3}
In the oscillator network discussed earlier, EE emerges as a noise-driven phenomenon and is often accompanied by synchronized spiking activity. A critical factor contributing to their formation is the presence of non-identical noise sources acting on individual oscillators. To elucidate this mechanism, we begin with a simplified case of two coupled oscillators ($N=$ 2), which serves as a minimal model to uncover how noise heterogeneity shapes the generation of EE. The two-neuron configuration is intentionally designed as a minimal framework to investigate the mechanistic influence of noise heterogeneity on the generation of EE. Specifically, this setup enables a detailed examination of how variations in the local noise standard deviations ($\sigma_1$ and $ \sigma_2$) interact with the global noise intensity ($D$) to shape the system’s collective dynamics. Such a reduced model facilitates a clear interpretation of the interplay between $\sigma$-values and $D$, which would otherwise be masked in larger ensembles due to mean-field averaging effects. If a slightly larger network (e.g., $N \ge$ 3) were considered, the same qualitative dynamical states would emerge, provided that $\sigma_i$ values are appropriately distributed within the range [0.01, 1]. However, in that case, the mean-field coupling would homogenize the network dynamics, making it difficult to discern the direct impact of local noise disparities. Therefore, the two-neuron system serves as the most transparent framework for isolating and understanding the fundamental role of noise variance in inducing and synchronizing EE.

\begin{figure}[h!]
\centering
\includegraphics[width=1.0\linewidth]{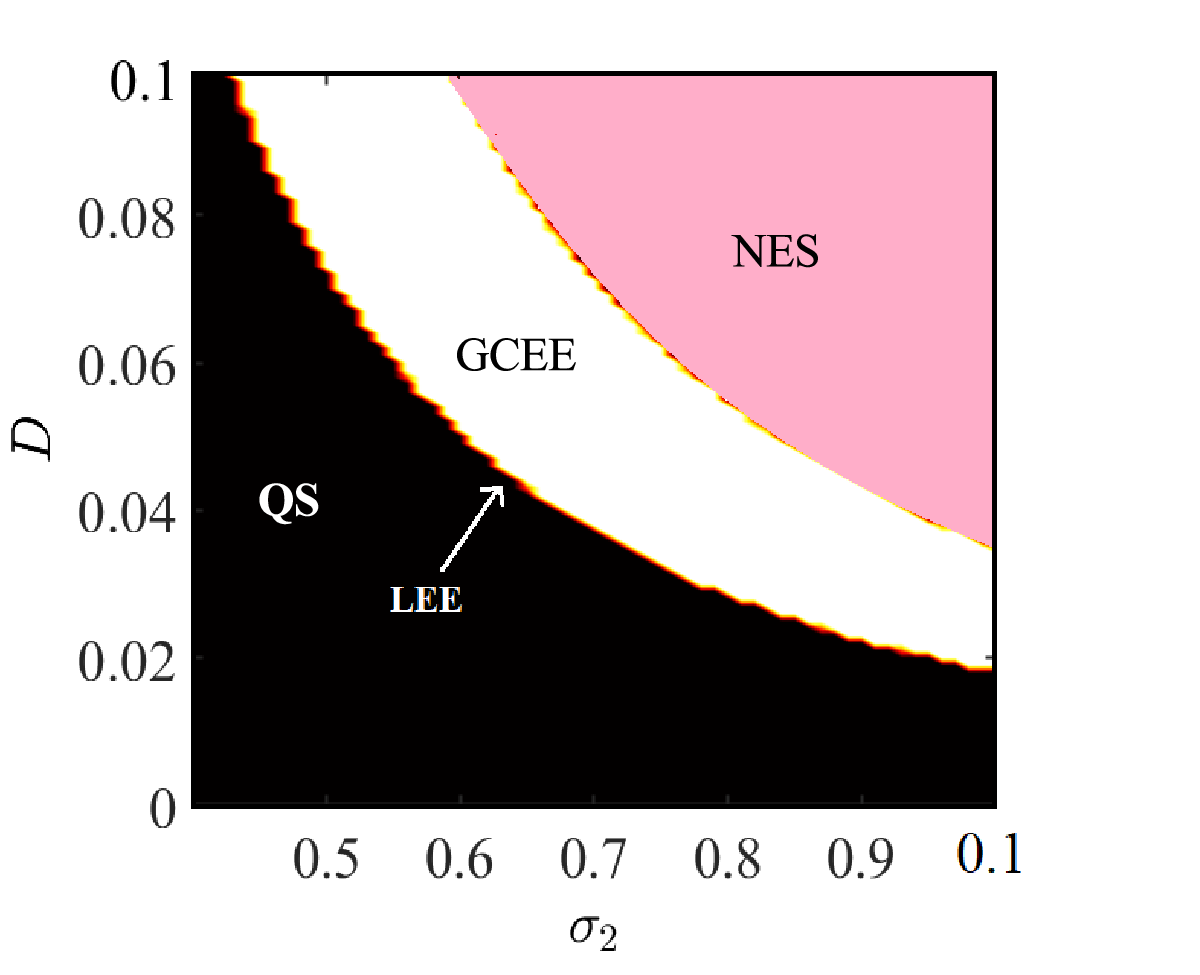}
\caption{\label{2sigd}Two-parameter phase diagram in the ($\sigma_2, D$) space for fixed $\sigma_1 = 0.01$ and $\kappa = 0.05$, illustrating the trade-off between noise heterogeneity and noise intensity in a two-oscillator system. The black region corresponds to the QS, while the white region denotes the GCEE regime, characterized by $Z > 0.8$. The light pink region represents the NES state, where frequent oscillations occur without meeting the EE threshold. The arrow highlights a narrow boundary corresponding to PCEE, with $Z \approx 0.5$, indicating partial participation of oscillators in EE activity.}
\end{figure}

To examine the role of noise disparity, we fix the standard deviation of the first oscillator’s noise at $\sigma_1 = 0.01$ and set the coupling strength to $\kappa = 0.05$. We then construct a two-parameter phase diagram in the ($\sigma_2, D$) plane, where a measure $Z$ quantifies the occurrence of EE (Fig.~\ref{2sigd}). The results reveal a clear trade-off between local standard deviations ($\sigma_1, \sigma_2$) and the global noise strength $D$. The black region corresponds to the QS state, where the system remains confined to low-amplitude oscillations. A narrow white band marks the emergence of the GCEE state, in which both oscillators undergo large synchronized excursions. Notably, the threshold for GCEE shifts toward lower values of $\sigma_2$ as $D$ increases, demonstrating that stronger noise can compensate for weaker heterogeneity. For example, at high noise intensities ($D \sim 0.06$), GCEE occurs even when $\sigma_2 \approx 0.5$. This observation is consistent with larger networks ($N=100$), where oscillators subjected to larger $\sigma$ values are more likely to spike, while those with smaller $\sigma$ values remain quiescent at low $D$ and moderate coupling. At higher $D$, however, even oscillators with smaller $\sigma$ values can spike, revealing an expanded susceptibility to EE. While a direct one-to-one correspondence between the two-oscillator case and large networks is not feasible due to system complexity, the underlying principle remains consistent: the emergence of EE is governed by the interplay between global noise intensity and local noise heterogeneity.

Within the GCEE band, a localized domain of PCEE is observed (highlighted by the arrow in Fig.~\ref{2sigd}), where the oscillators exhibit partial synchronization, moderate values of $Z$, and occasional extreme excursions. As both $D$ and $\sigma_2$ increase further, the system transitions into the NES regime (pink region), where noise dominates the dynamics, driving frequent large-amplitude oscillations without global synchrony.

\begin{figure}[h!]
\centering
\includegraphics[width=1.0\linewidth]{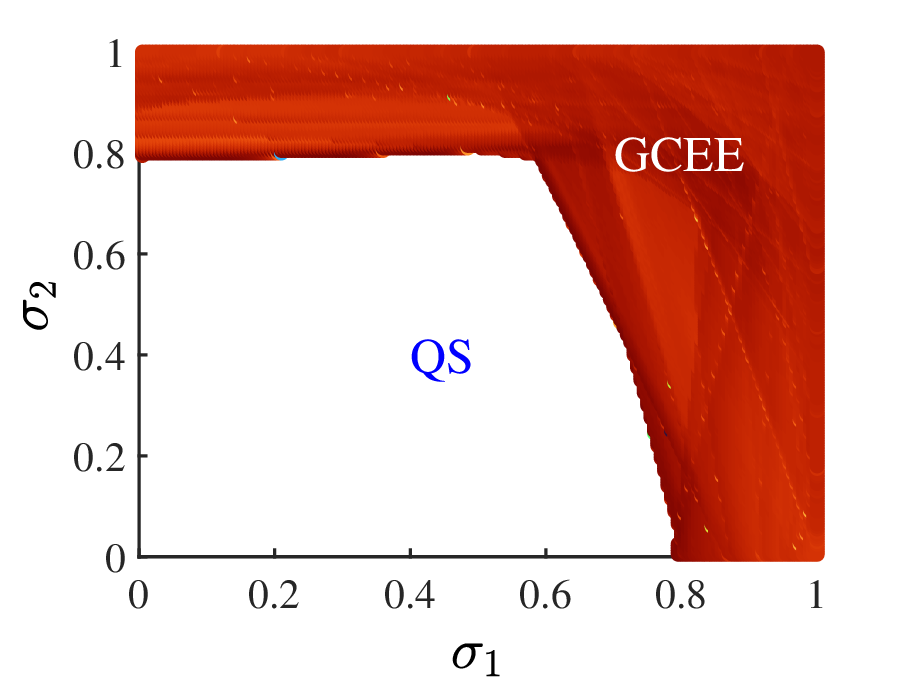}
\caption{\label{2sig}Two-parameter phase diagram in the ($\sigma_1, \sigma_2$) space illustrating the effect of noise heterogeneity for fixed values of $D = 0.03$ and $\kappa = 0.1$. The white region corresponds to the QS, where no spiking activity occurs, while the red region denotes the GCEE state, characterized by network-wide synchronized excursions.}
\end{figure}

To further probe the role of heterogeneity, we construct a two-parameter diagram in the ($\sigma_1, \sigma_2$) space at fixed coupling $\kappa=$ 0.1 and $D =$ 0.03, using a spike-based order parameter (Fig.~\ref{2sig}). This analysis reveals the presence of only two regimes: QS and GCEE. When both $\sigma_1$ and $\sigma_2$ are small, the noise distribution is narrow, and the system remains in the QS state. In contrast, when either $\sigma_1$ or $\sigma_2$ is significantly larger, EE are induced through mutual interactions, leading to GCEE. For instance, a large $\sigma_2$ can induce spiking in oscillator 1, and vice versa. This demonstrates that heterogeneity in noise strength, particularly at intermediate $D$, can act as a robust catalyst for EE generation. Interestingly, no NES regime is observed for this parameter set, as the chosen values lie within the transition domain between QS and GCEE.

Overall, these findings establish that the emergence of EE is primarily governed by heterogeneity in the standard deviation of noise inputs, modulated by global noise intensity. Strong coupling facilitates the spread of localized activity, but heterogeneous noise sources determine whether the system transitions from quiescence to EE regimes. This mechanism remains consistent across both small ($N = 2$) and large ($N = 100$) networks, highlighting the universal role of noise diversity in driving extreme collective dynamics.

\section{Discussion and conclusion}
\label{sec4}
In this work, we investigated the emergence of noise-induced EE in a globally coupled network of FHN oscillators subjected to heterogeneous noise sources. By systematically varying noise intensity and coupling strength, we identified five distinct dynamical regimes: (i) QS, where no large excursions occur; (ii) LEE state, restricted to a few oscillators; (iii) PCEE state, involving a significant fraction of oscillators; (iv) GCEE state, characterized by synchronized large-amplitude excursions across the entire network; and (v) NES state, where frequent oscillations occur without meeting the threshold for EE.

To distinguish among these states, we introduced a novel quantitative measure that captures the spatial extent of EE activity, complemented by a criterion based on the global mean field. This framework enabled a robust classification of local, partial, and global EE regimes. Furthermore, by employing a spike-based order parameter derived from Hilbert-phase analysis, we quantified synchronization during EE. We demonstrated that noise intensity and coupling strength jointly govern the transition from partial to globally synchronized extremes.

Our mechanistic analysis of a reduced two-oscillator system confirmed that heterogeneity in noise variance is a key driver of EE onset. Specifically, stronger noise fluctuations in one unit can induce extreme excursions in its partner through coupling, thereby promoting GCEE. This finding scales naturally to larger ensembles, where diversity in noise inputs interacts with coupling to determine the transition between different EE states.
Importantly, we demonstrate for the first time that heterogeneous noise alone can synchronize EE across a network--establishing noise not merely as a trigger but as an organizing principle for EE in excitable systems.
Importantly, we demonstrated for the first time that heterogeneous noise alone can synchronize extremes across a network, establishing noise not just as a trigger but as a robust organizing principle for EE in excitable systems.

These results extend the understanding of noise-induced behavior in coupled excitable systems, complementing earlier works that focused primarily on deterministic mechanisms of EE generation~\cite{feudel2013fhn,feudel2014fhn,feudel2017delay}. Although the FHN model used here is a simplified excitable framework with limited biophysical detail, it remains a robust representative for exploring complex neuronal interactions. Noise-driven transitions in neural networks have been widely reported in the context of resonance phenomena, where entrainment enhances signal processing. However, the study of noise-induced EE in neuronal populations remains relatively underexplored, and the present work contributes a significant step toward addressing this gap.

Additionally, our findings from the heterogeneous FHN network draw a conceptual parallel to experimental and computational studies on noise-induced epileptiform activity. Puzerey \textit{et al.} (2014) demonstrated that enhanced serotonergic signaling increases synaptic variability, facilitating the emergence of epileptiform oscillations \cite{puzerey2014elevated}. Similarly, Naze \textit{et al.} (2015) showed that even networks lacking intrinsic epileptogenic properties can exhibit seizure-like events when subjected to exogenous stochastic fluctuations, with network topology and heterogeneity shaping the resulting dynamics~\cite{naze2015computational}. Complementary in vivo studies by Luna-Munguía \textit{et al.} (2017) revealed that increasing synaptic noise through localized KCl microinjections elevates seizure probability, emphasizing the causal role of endogenous fluctuations in ictogenesis \cite{luna2017control}. In our theoretical framework, the non-identical noise sources serve as an analogue for both biological synaptic noise and external stochastic drives. Accordingly, we observe rare spikes at low noise, extreme events at intermediate noise, and frequent spiking at high noise, closely mirroring the qualitative dynamics reported in these experimental studies.

This correspondence suggests that noise-induced transitions in heterogeneous networks may constitute a unifying mechanism underlying the initiation and propagation of epileptiform activity across biological and computational systems. Our findings, therefore, hold broader implications for understanding pathological brain activity, such as seizure-like bursts in epileptic networks~\cite{gerster2020,cornejo2024}, and for developing predictive or control strategies in other complex systems, including power grids, climate models, and engineered oscillator networks.

In summary, this study advances our understanding of how stochasticity, heterogeneity, and coupling collectively contribute to the emergence of extreme network dynamics. By uncovering the conditions under which partially and globally synchronized EE emerge under heterogeneous noise, it opens new pathways for research on the prediction, control, and suppression of EE in neuronal and other nonlinear networked systems. Future work exploring multilayer and time-delayed architectures could provide deeper insights into the mechanisms that govern such sporadic, noise-driven phenomena.

\section*{Acknowledgments}
The research contributions of S.H., R.S., and V.K.C. are part of a project funded by the SERB-CRG (Grant No. CRG/2022/004784). The authors gratefully acknowledge the Department of Science and Technology (DST), New Delhi, for providing computational facilities through the DST-FIST program under project number SR/FST/PS-1/2020/135, awarded to the Department of Physics. The constructive feedback from anonymous reviewers is sincerely appreciated, as it significantly contributed to improving the clarity and quality of this manuscript.

\section*{Data Availability}
The data that support the findings of this study are available from the corresponding author upon reasonable request.

\bibliographystyle{elsarticle-num} 
\bibliography{cas-refs}

\end{document}